\documentclass[12pt]{article}
\input epsf.sty
\newcommand{\ra}{\rangle}
\newcommand{\la}{\langle}
\newcommand{\T}{\chi_{T}(k)}
\newcommand{\Tm}{\chi_{T}(k')}
\newcommand{\Cn}{{\cal C}_n}

\newcommand{\CC}{{\cal C}}
\newcommand{\OO}{{\cal O}}
\newcommand{\QQ}{{\cal Q}}

\newcommand{\EE}{{\cal E}}

\newcommand{\wh}{\widehat}

\newcommand{\NN}{{\cal N}}
\newcommand{\TT}{{\cal T}}

\newcommand{\be}{\begin{equation}}
\newcommand{\ee}{\end{equation}}
\newcommand{\ben}{\begin{eqnarray}\displaystyle}
\newcommand{\een}{\end{eqnarray}}
\newcommand{\refb}[1]{(\ref{#1})}
\newcommand{\p}{\partial}
\newcommand{\sectiono}[1]{\section{#1}\setcounter{equation}{0}}

\begin{document}
{}~
\hfill\vbox{\hbox{hep-th/0111153}\hbox{CTP-MIT-3210}
\hbox{CGPG-01/11-2}
\hbox{PUPT-2012} 
}\break

\vskip .6cm

\centerline{\large \bf A Note on a Proposal for the
Tachyon State}

\medskip

\centerline{\large \bf in Vacuum String Field  Theory}

\vspace*{4.0ex}

\centerline{\large \rm Leonardo Rastelli$^a$,
Ashoke Sen$^b$ and Barton Zwiebach$^c$}

\vspace*{4.0ex}

\centerline{\large \it ~$^a$Department of Physics }

\centerline{\large \it Princeton University, Princeton, NJ 08540,
USA}

\centerline{E-mail:
        rastelli@feynman.princeton.edu}

\vspace*{2ex}

\centerline{\large \it ~$^b$Harish-Chandra Research
Institute}

\centerline{\large \it  Chhatnag Road, Jhusi,
Allahabad 211019, INDIA}

\centerline {and}
\centerline{\large \it Department of Physics, Penn State University}

\centerline{\large \it University Park,
PA 16802, USA
}

\centerline{E-mail: asen@thwgs.cern.ch, sen@mri.ernet.in}

\vspace*{2ex}

\centerline{\large \it $^c$Center for Theoretical Physics}

\centerline{\large \it
Massachussetts Institute of Technology,}

\centerline{\large \it Cambridge,
MA 02139, USA}

\centerline{E-mail: zwiebach@mitlns.mit.edu}

\vspace*{5.0ex}

\centerline{\bf Abstract} 
\bigskip 
We discuss the proposal of Hata and
Kawano for the tachyon fluctuation around a solution of vacuum string
field theory representing  a D25 brane. 
We give a conformal field theory construction of their state -- 
a local insertion of a tachyon vertex operator on the sliver 
surface state, and explain  why the on-shell momentum condition emerges
correctly.  We also show that a  naive computation of the D25-brane 
tension   using data for the three point coupling of this state
gives an answer that is $(\pi^2/3)(16/27\ln 2)^3 \simeq 2.0558$ times the
expected answer.  We demonstrate that this problem arises because 
the HK state does not satisfy the equations of motion in a strong sense 
required for the computation of D-brane tension  from the on-shell 3-tachyon
coupling.

\vfill \eject

\baselineskip=16pt

\tableofcontents

\sectiono{Introduction}\label{s0}

In a very stimulating paper \cite{0108150} Hata and Kawano identified a
particular state (hereafter refered to as the HK state) as an off-shell
tachyon on the D25-brane solution in vacuum string field theory
(VSFT) \cite{0012251, 
0106010, 0111129}, 
and showed that the linearized
field equations around the D25-brane background lead to the correct
on-shell condition for the tachyon.  However, when they computed the
three tachyon amplitude using this on-shell state, and fixed the
normalization of the action by requiring this amplitude to coincide
with the known answer, the energy density $\EE_c$ associated with the
D25-brane solution did not agree with  the
expected answer $\TT_{25}$ for the D25-brane tension.  A refinement of
their
analysis has very recently appeared in \cite{0111034}.
In this analysis the result for
the $\EE_c/\TT_{25}$ ratio
still did not agree
with the expected answer for the D25-brane tension, but came out to be
about twice the expected value. This suggested the possibility
that the perhaps the sliver would represent a state with two D-branes.  

In both these papers the solution was constructed using the oscillator
representation of the states. Although much of the calculation can be
carried out analytically,
the final expression for the tachyon state involves
inverses and square roots of infinite dimensional matrices.
The expression for the $\EE_c/\TT_{25}$ ratio also
involves determinants of infinite dimensional matrices.
Thus the final answers for these quantities have to be
computed numerically by truncating the infinite dimensional matrices to
finite dimensional ones.

In this note we propose a description of the HK state using the language
of boundary conformal field theory (BCFT) which allows us to give 
an explicit simple analytic form 
for the tachyon state as well as a closed form 
 for the
$\EE_c/\TT_{25}$ ratio calculated in \cite{0111034}. In particular we
show that the $\EE_c/\TT_{25}$ ratio computed in ref.\cite{0111034} is
given by:
\be \label{efirst}
{\EE_c \over \TT_{25}} = {\pi^2 \over 3} \bigg( {16\over 27 \ln
2}\bigg)^3
\simeq 2.0558\,.
\ee
This rules out the two D-brane interpretation of the sliver, and thus
appears to be problematic. Nevertheless, we explain that 
a problem with the proposed tachyon state invalidates 
the computation of the
$\EE_c/\TT_{25}$ ratio in this framework. Thus our conclusion is that the
computation performed 
in refs.\cite{0108150,0111034} is not expected to
produce the correct expression for the $\EE_c/\TT_{25}$ ratio. In a 
nutshell, the field equation for the on-shell tachyon state
is a state  
whose inner product with Fock space states vanishes, but whose
inner product with another tachyon state does not vanish.  Thus
the field equation does not hold in a strong sense for this
tachyon state. The question of finding a fully consistent
tachyon state thus remains open. 

The paper is organized as follows. In section \ref{s2} we propose a
conformal field theory description of the HK state, show that it yields
the correct on-shell condition, and give numerical evidence that it
corresponds to the same state discussed in refs.\cite{0108150,0111034}. In
section \ref{s3} we analytically compute the $\EE_c/\TT_{25}$ 
ratio by 
naively following the procedure outlined in \cite{0108150,0111034} and
arrive at \refb{efirst}. We then show that one of the steps in this
calculation is illegal, and hence \refb{efirst} is not a valid result.

\sectiono{Conformal field theory description of the HK state} \label{s2}

In this section we shall give a conformal field theory
description of the
HK state of ref.\cite{0108150}, and show analytically that
if we take the inner product of
the linearized equations of motion around
the D25-brane background with a Fock space state, we are led to the
correct mass-shell condition for the tachyon on the D25-brane.
Thus this
gives an analytic proof of the result of ref.\cite{0108150}.

We shall proceed in two steps.
\begin{enumerate}
\item First we shall propose a conformal field theory
construction of the
HK state and show that it gives the correct on-shell condition when we
take the inner product of the equations of motion with the Fock space
state.
\item Then we present numerical evidence that our proposal for the HK
state agrees with the state constructed in \cite{0108150}.
\end{enumerate}

\subsection{BCFT construction of the HK state}

The VSFT action is given by~\cite{0106010}
\be \label{ehk1}
S = - 
\Big[{1\over 2} \langle \Psi| \QQ |\Psi\rangle +
{1\over 3}
\langle \Psi| \Psi * \Psi\rangle \Big]
\ee
where $|\Psi\rangle$ is the string field represented by a ghost number one
state in the matter-ghost boundary conformal field theory (BCFT),
$\langle A| B\rangle$ denotes BPZ inner product, $*$ denotes star
product~\cite{OSFT}
and
$\QQ$ is an operator of ghost number one made purely of ghost fields.
We have normalized $|\Psi\rangle$ and $\QQ$ such that any overall
normalization constant multiplying the action, involving the open string
coupling constant, is  absorbed into a redefinition of $|\Psi\rangle$ and
$\QQ$. 
For definiteness we shall take the matter part of
BCFT to be the one associated with a D25-brane.
The
equations of motion are
\be \label{ehk2}
\QQ|\Psi\rangle + |\Psi * \Psi\rangle = 0\, .
\ee
As usual we look for factorized solution of the equations of motion:
\be \label{ehk3}
|\Psi\rangle = |\Psi_g\rangle \otimes |\Psi_m\rangle\, ,
\ee
where $|\Psi_g\rangle$ and $|\Psi_m\rangle$ are the ghost and the matter
parts of the solution, satisfying,
\be \label{ehk4}
\QQ|\Psi_g\rangle + |\Psi_g * \Psi_g\rangle = 0\, ,
\ee
and
\be \label{ehk5}
|\Psi_m\rangle = |\Psi_m*\Psi_m\rangle\, .
\ee
The solution $|\Psi_g\ra$ is taken to be universal,  $-$ the same for all 
D-brane solutions. On the other hand, $|\Psi_m\ra$ 
corresponding to different D-brane solutions differ from each other.

A solution to the matter part of the equation of motion, describing the
D25-brane solution, is provided by the
matter sliver,
defined through the relation~\cite{0006240, 0105168,0106010} 
\be \label{edefsl}
\la \Xi_m |\psi\ra = \lim_{n\to\infty} \, 
\NN \, \la f\circ\psi(0)\ra_{\Cn}
\ee
where $f(z)=\tan^{-1}z$, $|\psi\rangle$ is a state in the matter
Hilbert space, $\NN$ is a normalization factor which ensures that
$\Xi_m$ squares to itself under $*$ multiplication, and $\la~\ra_{\Cn}$
denotes correlation function of the matter BCFT
on a semi-infinite
cylinder $\Cn$ of circumference $n\pi/2$ obtained by making the
identification $\Re(z)\simeq \Re(z)+n\pi/2$ in the upper half plane. In
the $n\to\infty$ limit $\CC_n$ approaches the upper half plane.
We have
\be \label{ehk6}
\langle \Xi_m|\Xi_m\rangle = K V^{(26)}\, ,
\ee
where
$V^{(26)}$ is the volume of the 26-dimensional space-time, and $K$ is a
normalization constant arising due to the conformal anomaly in the matter
sector. Here we are using the 
normalization convention:
\be \label{ehk7}
\langle k| k'\rangle = (2\pi)^{26} \delta(k+k')\, ,
\ee
where $|k\rangle$ denotes the Fock vacuum with momentum $k$.
We also have the identification $(2\pi)^{26} \delta(0)=V^{(26)}$.

We now propose that the HK tachyon state is given by
\be \label{eHK}
|\Psi_g\ra \otimes |\T\ra  
\ee
where $\Psi_g$, satisfying \refb{ehk4}, is the same solution that appears
in
the construction of the classical solution describing a D-brane
configuration, and
\be \label{T}
\la \T |  \, \psi \ra \equiv \NN \lim_{n \to \infty} n^{2 k^2} \,
\la      e^{i k\cdot X(n \pi/4)}    f\circ \psi(0)     \ra_{\Cn}
\quad
\forall \psi \,.
\ee
In other words we simply insert a tachyon vertex operator in the
middle of the (matter) sliver, at the diametrically opposite
 point with respect to the 
puncture. (Recall that in our notations the cylinder $\Cn$  
defined as $-{\pi\over 4}\leq \Re (z) \leq {n\pi\over 2} - {\pi\over 4}$, 
$\Im (z) >0$. It has
circumference $n
\pi/2$ and the puncture is at $f(0)=0$. The local coordinate patch
extends between the vertical lines  $\Re(z) = \pm {\pi\over 4}$,
$\Im(z) >0$, which represent the two halves of the open string).
The explicit factor of $n^{2k^2}$ ensures that the state has finite
overlap
with
Fock
states except for the overall normalization factor $\NN$ which could be
infinite. The normalization factor of $\NN$ has been included for
convenience, so that the overall normalization in any formula involving
$\T$ can be fixed by noting that for $k=0$ we must recover the
corresponding formula for $\Xi_m$.

Representing the state $|\T\ra$ and the sliver in terms of correlation 
function on $\CC_n$ has the advantage that the $*$-product / BPZ inner 
product of two such states has a simple geometric 
interpretation~\cite{0105168,0106010}. We 
first cut 
open each cylinder $\CC_n$ along the lines $\Re(z)=\pi/4, (2n-1)\pi/4$, 
representing the left and the right halves of the string respectively, to 
get a strip of width $(n-1)\pi/2$. We then glue the right end of the first 
strip to the left end of the second strip to get a bigger strip of width 
$(n-1)\pi$. In the case of BPZ inner product, we simply glue the left and 
the right ends of the resulting strip to get a cylinder $\CC_{2n-2}$ of 
circumference $(n-1)\pi$. In the case of $*$-product we glue 
another strip of width $\pi/2$, representing the local coordinate patch, 
to one end of the big strip, and then glue the left and the right ends of 
the resulting strip to get a cylinder of width $(2n-1)\pi/2$. 
The final result is then expressed as a correlation function on this big 
cylinder. 
For details 
of this construction, see ref. \cite{0105168}.

In the $n\to\infty$ limit, 
the two operators $e^{ik\cdot X(n\pi/4)}$ and
$f\circ\psi(0)$ in \refb{T} are infinite distance apart, and as a result 
the higher the dimension of the operator in $f\circ\psi(0)$ the
more supressed is its contribution to the correlator. 
In fact, due to the compensating factor $n^{2k^2}$, we will see that 
only the lowest dimension operator in $f\circ\psi(0)$ carrying momentum $-k$
will contribute to the
correlator.
More precisely, we can write
\be
f \circ \psi (0) = a_\psi e^{-i k\cdot X(0)} + {\rm descendents \,of  \,}
e^{-i k\cdot X(0)} +
{\rm other \,conformal \, blocks}\, ,
\ee
where the coefficients $a_\psi$ depends only on $\psi$.
The two-point function (\ref{T}) will be non-zero
only if $ a_\psi \neq 0$. Indeed the primary $e^{-ik\cdot X(0)}$
is the only primary non-orthogonal to the insertion $e^{i k\cdot X}$.
Thus the other conformal blocks cannot contribute. 
Moreover,
descendents of  $e^{-ik\cdot X(0)}$, which give a non-vanishing
two point function for finite $n$,  can be ignored since
they have higher
dimension
and their contribution goes to zero for $n$ large. Thus
we have  
\ben
\label{cvb}
\la \T |  \, \psi \ra &=& \lim_{n\to\infty}\Big[ \NN \,a_\psi \, n^{2 k^2}
\la      e^{i k\cdot X(n \pi/4)}    e^{-i k\cdot X(0)}     \ra_{\Cn}\Big]
\nonumber \\
&=& \lim_{n\to\infty} \Big[\NN\, a_\psi \,  n^{2 k^2} \,
\bigg({4\over n}\bigg)^{2k^2} 
\la e^{i k X(-1)} e^{-i k X(1)} \ra_{D}\Big]\,,
\een
where we have used the change of variables $w =\exp(4 i z_n/n)$
to map the cylinder $\Cn$ parametrized by $z_n$ to the unit disk $D$
parametrized by $w$. 
We now record that\footnote{For the two point function of operators on the
boundary of the unit disk we have $\la X(z) X(w)\ra = -{1\over 2} \{ 
\ln |z-w|^2 + \ln |z - {1\over \bar w} |^2 \}$.}
\ben \label{egenfor}  
\la e^{i k\cdot X(e^{i\theta})}    e^{i k'\cdot X(e^{i\theta'})} 
\ra_{D}  
&=& \Big[ 2 \sin\Bigl({\theta - \theta'\over 2}\Bigr) \Big]^{2 k\cdot k'} 
(2\pi)^{26} \delta(k+k')\, \nonumber\\
& =& 
[d(\theta,\theta')]^{2 k\cdot k'} 
(2\pi)^{26} \delta(k+k')\,,
\een
where $d(\theta,\theta')$ is the distance between the points
$\exp(i\theta)$ and $\exp(i\theta')$. This result gives 
\be
\la e^{i k\cdot X(-1)}    e^{-i k\cdot X(1)}     \ra_{D}  =
\frac{1}{2^{2k^2}}
V^{(26)}\,,
\ee
and therefore, back in \refb{cvb} we get  
\be \label{T=}
\la \T |  \, \psi \ra = 2^{2k^2} \,\NN \, a_\psi \, V^{(26)}\,.
\ee

\bigskip
As discussed above,  
we can compute the star products
$\Xi_m*\T$ and $\T*\Xi_m$ by method similar
to the ones described in \cite{0105168,0106010}.  
In doing this we keep $n$ 
large and fixed and take the $n\to\infty$ limit at the end of the
computation.
We get, for large $n$:
\ben
&& \Bigl\langle \Xi_m * \T +\T * \Xi_m \, | \psi \ra   \\
&=&
\NN\, a_\psi \, n^{2k^2}  \Big[ \la      e^{i k\cdot X(n \pi/4)}   e^{-i k
X(0)}
\ra_{{\cal
C}_{2n-1} } +  \la      e^{i k\cdot X((3n-2) \pi/4)}   e^{-i k
X(0)}
\ra_{{\cal
C}_{2n-1} } \Big] \nonumber \\
&=&
\NN\, a_\psi \,  n^{2 k^2} \,  \bigg( {4\over 2n-1}\bigg)^{2k^2}
\Big[ \la e^{i k\cdot X(i)}
e^{-i k
X(1)}
\ra_{D} + \la e^{i k\cdot X(-i)}    e^{-i k
X(1)}
\ra_{D}\Big]\,, \nonumber 
\een
where we have used $w=\exp(4 i z_{2n-1}/(2n-1))$ to map the cylinder
$\CC_{2n-1}$
to the unit disk. 
Using eq.\refb{egenfor}
we get, for large $n$, 
\be \label{Tsliver=}
\la \Xi * \T +\T * \Xi \, | \psi \ra  = 2^{k^2+1} \, \NN\, a_\psi  \,
V^{(26)}\, ,
\ee
This is precisely the same as (\ref{T=}) if
\be \label{eons}
k^2 = 1\, .
\ee
Thus if \refb{eons} is satisfied, $\T$ satisfies the
linearized equations
of motion:
\be \label{eleom}
\la \T |\psi\ra = \la (\T*\Xi_m +\Xi_m * \T)|\psi\ra\,,
\ee
for any Fock space state $|\psi\rangle$.
\refb{eons} gives the correct on-shell condition for the tachyon living
on the D-brane.

Finally we note that we can replace the vertex operator $e^{ik\cdot X}$ in
eq.\refb{T} by any primary operator, and an 
analysis identical to the
one carried out here will show that the corresponding state satisfies the
linearized equations of motion if the primary operator has dimension one.

\subsection{Comparison with the algebraic description of the HK state}

Now we would like to find the oscillator form of the state $\T$, and
compare this with the HK state constructed in ref.\cite{0108150}.
To
this end, let us derive current conservation laws
for $|\T\ra$.  
Consider a scalar function $F(z)$ on $\Cn$ (more precisely on the double
cover of $\Cn$ obtained by removing the restriction $\Im(z_n)\ge 0$)
that is regular everywhere
except for possible poles at $z=0$.  Then we claim that
\be \label{mothercons}
\la \T |   \, \oint_{\cal C} d\xi \, F(f(\xi)) \,
\p X_\mu(\xi) =  i k_\mu \, F_0\,  \la \T | \,,  
\ee
where $F_0 \equiv F(n\pi/4)$, $f(\xi)=\tan^{-1}\xi$, and
${\cal C}$ is a contour in the $\xi$ plane
surrounding
the origin 
in the usual counterclockwise direction. This can be seen as follows.
The inner product of the left hand side of eq.\refb{mothercons} with a
Fock space state $|\phi\rangle$ is given by:
\ben \label{eyy1}
&&\NN\, \lim_{n\to\infty} n^{2k^2} \langle e^{ik\cdot X(n\pi/4)} \ointop_C
d\xi
F(f(\xi)) f\circ \p X(\xi) f\circ \phi(0)\rangle_{\Cn} \nonumber \\
&=& \NN\, \lim_{n\to\infty} n^{2k^2} \langle e^{ik\cdot X(n\pi/4)} \ointop
dz
F(z)
\partial X(z)  f\circ \phi(0)\rangle_{\Cn} \, ,
\een
where $z=f(\xi)$. Deforming the $z$ contour on $\Cn$, 
since the function
$F$ has no singularity at infinity, we get just a contour surrounding
the point $z= n\pi/4$ in the clockwise direction.  Using the OPE
\be \label{ope}
\p X_\mu(z) \, e^{i k\cdot X(w)} \sim -\frac{i k_\mu}{z-w} e^{i k\cdot
X(w)}\,,
\ee
we see that we pick up a pole equal to $+i F_0 k_\mu    e^{i k\cdot
X(n\pi/4)}$. This gives eq.\refb{mothercons}. Thus
the effect of the tachyon insertion is simply
to add an inhomeogenous term to the
current conservation laws for the sliver
surface state.
Since $\Cn$ approaches the upper half plane as $n\to\infty$, in this
limit we can take
$F(z)$ to be a function on upper half plane with possible poles at the
origin, and $F_0$ to be the value of $F(z)$ at $\infty$.

For $F(z)=1$, (\ref{mothercons})
gives the expected conservation law for the momentum
$a_0$:\footnote{As in \cite{0108150}, 
$i\sqrt{2}\partial  
X(\xi) \equiv
a_0 \xi^{-1} +\sum_{n\ne 0}
\sqrt{|n|} a_n\, \xi^{-n-1} $, with the BPZ conjugate of $a_n$
being $(-1)^{n+1} a_{-n} = (-1)^{n+1} a_n^\dagger$. 
In this convention, the 
operator product of two $\p X$ in the bulk is given by $\p X^\mu(z) \p 
X^\mu(w) 
\simeq - \eta^{\mu\nu} / 2(z-w)^2$, and that between a $\p X$ in the bulk 
and 
$e^{ik\cdot X}$ on the boundary is given by eq.\refb{ope}.} 
\be \label{emomcons}
\langle \T| a_0^\mu = - \sqrt 2 k^\mu \langle \T|\, .
\ee
Using this result, and \refb{mothercons} we can write a convenient form of
the general conservation law:
\be \label{fatherconst}
\langle \T|\Bigl( -F(\infty) a_0^\mu + \oint_0 d\xi \,F (\tan^{-1}(\xi))
\,i\sqrt 2
\partial X^\mu (\xi) \Bigr) = 0\,,
\ee
where $F(z)$ is regular except for poles at $z=0$. We can derive
conservation laws for higher $a_{n}^\dagger$'s by taking
$F(z)$
to have a
pole
of order $n$* at $z=0$ and converting to the local coordinate $\xi$ using
$z =\tan^{-1}\xi$. For example, taking
$F(z) = 1/z^2$, we have $F(\infty)=0$ and   
we obtain
\be
\la \T |   \, \oint_{\cal C} d \xi \,
\frac{1}{(\tan^{-1}\xi)^2}
\,i\sqrt{2}\, \p X_\mu(  
\xi) =  0\,,
\ee
which gives
\be
(a_{2} + \frac{\sqrt{2}}{3} a_0 -\frac{1}{15} a_{-2} + \frac{32
\sqrt{2}}{945}
a_{-4}
+ \dots)
 | \T \rangle = 0\,.
\ee
By choosing $F(z)=F_n(z)$ such that
\be
\label{neq}
\sqrt{n} F_n(\tan^{-1}\xi)=\xi^{-n}+\OO(\xi^2)\,,
\ee
the conservation laws \refb{fatherconst}
can be written in the form
\be \label{HKcons}
(a_n +  \sum_{m=1}^\infty   S_{nm} a^\dagger_m
+ \hat t_n  a_0) \, |\T\ra
= 0
\, ,
\ee
where $S_{nm}$ is guaranteed to be the matrix
that appears in the oscillator representation
of the matter sliver,  
\be \label{msli}
\Xi_m = {\cal N} \exp\Bigl(-{1\over 2}
a^\dagger \cdot S \cdot a^\dagger\Bigr) | 0 \rangle\, .
\ee
Moreover, it is clear from \refb{neq} that the integral
term in \refb{fatherconst} gives no contribution to $\hat t_n$
and therefore we have 
\be
\label{tform}
\hat t_n = - (-1)^n F_n(\infty) \,.
\ee
Twist invariance of $\T$ imples that the momentum
coefficients $\hat t_n$ vanish for odd $n$.

The functions $F_n$ are readily constructed using \refb{neq} and
power series expansions.  For the first six nonvanishing cases we find
\ben  
\sqrt{2}F_2(z)& =& \frac{1}{z^2} - \frac{2}{3}  \,, \\
\sqrt{4} F_4(z)& =& \frac{1}{z^4}- \frac{4}{3 z^2}+ \frac{26}{45}\,,\nonumber
\\ \sqrt{6}F_6(z) & =& \frac{1}{z^6}-\frac{2}{z^4} + \frac{23}{15 z^2} -
\frac{502}{945}  \,,\nonumber \\
\sqrt{8}F_8(z)
& =& \frac{1}{z^8} -\frac{8}{3 z^6}+ \frac{44}{15 z^4}-\frac{176}{105
z^2}+
\frac{7102}{14175}\,  , \nonumber \\
\sqrt{10}F_{10}(z) & =& \frac{1}{z^{10}}
-\frac{10}{3 z^8}  + \frac{43}{9 z^6}-\frac{718}{189 z^4}+\frac{563}{315
    z^2}-\frac{44834}{93555} \,, \nonumber \\
\sqrt{12} F_{12}(z) & =& \frac{1}{z^{12}} - \frac{4}{z^{10}}
+ \frac{106}{15 z^8} -
    \frac{1360}{189 z^6} + \frac{21757}{4725 z^4} -
    \frac{6508}{3465 z^2} + \frac{295272982}{638512875} \, . \nonumber
\een
Using \refb{tform} we now simply read  
\ben
&& \hat t_2 =\frac{\sqrt{2}}{3} \cong 0.47140,   \; \; \quad \quad \quad
\hat
t_4 =-\frac{13}{45} \cong
-0.28889, \quad \\
&& \hat t_6 =\frac{502}{945 \sqrt{6}} \cong 0.21687,   \; \quad \quad
 \hat t_8 = - \frac{ 3551}{14175 \sqrt{2}} \cong -0.17714, \nonumber \\
&& \hat t_{10} = \frac{22417
\sqrt{2}}{93555 \sqrt{5}}\cong 0.15154,
 \quad \hat t_{12} =-\frac{147636491}{638512875 \sqrt{3}}  \cong
-0.13349\,.
\nonumber
\een
The conservation laws (\ref{HKcons})
lead to the following oscillator form of $\T$:
\be \label{HKlike}
|\T\ra   
\sim \exp \left( - a_0 \cdot \hat t \cdot a^\dagger -
\frac{1}{2}
a^\dagger
\cdot S \cdot a^\dagger \right) | k \ra \, .
\ee
The HK state of ref.\cite{0108150} has 
exactly the same form, 
where the coefficients $t_n$, labelled without a hat to distinguish
them from the predicted coefficiens $\hat t_n$,  were
computed in 
in terms of the matter sector
Neumann coefficients of the 3-string vertex. We have 
evaluated the $t_n$'s in
level truncation by 
using  $L \times L$ matrices and  the numerical
results are shown in table
\ref{t3}. We find good evidence that $\Psi_g \otimes \T$
coincides with the HK state.

\begin{table}
\begin{center}\def\st{\vrule height 3ex width 0ex}
\begin{tabular}{ |l|  l|  l|  l|  l|  l| l |               } \hline
$L$ &   $t_2  $ &   $  t_4  $   & $ t_6 $  &
$t_8   $ & $  t_{10}  $    &   $ t_{12} $
\st  \\[1ex]
\hline
\hline
100 &      0.43097    &    -0.25050       &    0.18064     &
-0.14271      &    0.11861     &  -0.10183
\st\\[1ex]
\hline
200 &    0.43734  &     -0.25656          &     0.18637  &    -0.14815
&  0.12381      &  -0.10682
\st\\[1ex]
\hline
300 & 0.44040 &     -0.25948         &     0.18914     &      -0.15079
&    0.12634        &  -0.10925
\st\\[1ex]
\hline
400 &    0.44234       &       -0.26133       &    0.19088  &
-0.15245     &  0.12793       &   -0.11079
\st\\[1ex]
\hline
500 &  0.44373  &       -0.26265     &      0.19213      &     -0.15364
&     0.12907        &  -0.11188
\st\\[1ex]
\hline
$\infty$ & 0.48018 &   -0.29736        &   0.22496       &  -0.18487
&   0.15896    & -0.14060
\st\\[1ex]
\hline
\end{tabular}
\end{center}
\caption{ Numerical results for $t_n$ at different level
approximation.
The last row
shows the interpolation of the various results
to $L=\infty$, obtained via a fitting function of the
form $a_0 + a_1/\ln(L)  $.} \label{t3}
\end{table}

\sectiono{Problems in the computation D25-brane tension}
\label{s3}

In this section we shall first perform a calculation similar to the ones
in
refs.\cite{0108150,0111034} to compute the D-brane tension and obtain an
answer that is
$\pi^2 (16 / 27\ln 2)^3 / 3 = 2.056$ times the expected answer.
We also show, however,  
that the on-shell HK state fails to satisfy the
linearized equations of motion when we take the inner product of the
equations of motion with the HK
state itself. This  
invalidates the calculation of the D-brane
tension.

We proceed in two steps.
\begin{enumerate}
\item First we show how a naive calculation of the D-brane tension yields
an answer that is too big.
\item We then show that the on-shell HK state 
fails to satisfy the equations of motion when we take its inner
product with an HK
state, and as a result the calculation in the first step breaks down.
\end{enumerate}

\subsection{Naive computation of the D25-brane tension}

In this computation we shall approximate the sliver as well as the HK
state $\T$ defined in the previous section by defining them in terms
of correlation
functions on a cylinder $\Cn$ with finite $n$, and take the $n\to\infty$
limit only at the end of the computation. We define the off-shell tachyon
field $T(k)$ in  momentum space through the expansion:
\be \label{ehk11}
|\Psi\rangle = |\Psi_g\rangle \otimes \Big(|\Xi_m\rangle + \int d^{26} k\,
n^{-k^2}  \,  T(k) \, |\T\ra
+ \ldots \Big)\,.
\ee
The $n^{-k^2}$ normalization factor has been included so that the
kinetic term for near on-shell tachyons will have no $n$ dependence.
With the help of
\refb{ehk4},
\refb{ehk5} the action
$S(|\Psi\rangle)$ computed for such a field configuration is given by:
\ben \label{ehk12}
S &=& S(|\Psi_g\rangle * |\Xi_m\rangle) \nonumber \\
&& - 
\langle \Psi_g | \QQ|\Psi_g\rangle \bigg[{1\over 2}
\int d^{26} k \, d^{26}k'\,\, T(k)\, T(k')\, n^{-k^2-k^{\prime2}} 
\\
&& ~ \qquad \qquad \qquad\times \la \chi_{T}(k')
|\Big(|\T\ra -
|\Xi_m * \T \ra - |\T * \Xi_m\ra \Big) \nonumber
\\
&& + {1\over 3} \int d^{26}k_1 d^{26}k_2 d^{26}k_3 \,\,T(k_1) T(k_2) T(k_3)
n^{-k_1^2-k_2^2-k_3^2} \la \chi_{T}(k_1) | \chi_{T}(k_2) *
 \chi_{T}(k_3)\ra\bigg] \,.\nonumber
\een
Now, from eqs.\refb{T=} and \refb{Tsliver=} we have
\be \label{ehk13}
|\Xi_m * \T \ra + |\T * \Xi_m\ra = 2^{k^2-1} |\T\ra \,.
\ee
Substituting this into \refb{ehk12} we get:
\ben \label{ehk14}
S &=& S(|\Psi_g\rangle * |\Psi_m\rangle)  \\
&& - 
\langle \Psi_g | \QQ|\Psi_g\rangle \bigg[{1\over 2}
\int d^{26} k d^{26}k'\,\, T(k) T(k') n^{-k^2-k^{\prime2}}
(1-2^{k^2-1}) \la \chi_{T}(k')
|\T\ra \nonumber \\
&& + {1\over 3} \int d^{26}k_1 d^{26}k_2 d^{26}k_3 \,\,T(k_1) T(k_2) T(k_3)
n^{-k_1^2-k_2^2-k_3^2} \la \chi_{T}(k_1)|  \chi_{T}(k_2)*
 \chi_{T}(k_3)\ra\bigg] \,. \nonumber 
\een 
We shall see in the next subsection that the use of \refb{ehk13} to obtain
\refb{ehk14}  is  illegal -- 
\refb{ehk13} holds in the weak sense that its inner product with any Fock
space state vanishes, but it does not hold when we take its inner product
with $\langle \chi_{T}(k')|$. Nevertheless we shall now proceed to analyze
the implications of \refb{ehk14}.

We shall analyze the quadratic and cubic terms separately for near
on-shell tachyon $k^2\simeq 1$.
The quadratic term for near on-shell tachyon takes the form:
\be \label{ehk16a}
S^{(2)} \simeq {1\over 2 
} \langle \Psi_g | \QQ|\Psi_g\rangle
\ln 2
\int d^{26} k \, d^{26} k' \,  n^{-k^2-k^{\prime 2}} (k^2-1) T(k) T(k')
\la \chi_{T}(k') | \chi_{T}(k)\ra \, .
\ee
The inner product is readily calculated
\ben \label{eorig} 
\la  \chi_{T}(k') |  \, \T \ra  
&=&  K \,   n^{2 k^2 + 2
k^{\prime 2}}
\la      e^{i k'\cdot X((n-1) \pi/2)}    e^{i k\cdot X(0)}
\ra_{C_{2n-2}} \nonumber \\
&=& K
\, n^{2 k^2 +2
k^{\prime 2} } \,  
\bigg({2\over n-1}\bigg)^{k^2 + k^{\prime 2}}
\la e^{i k'\cdot
X(-1)} e^{i k \cdot X(1)} \ra_{D} \nonumber \\
&\simeq& (2\pi)^{26} \, K\, \delta(k+k') \, n^{2k^2}\, .
\een
The overall normalization has been fixed so that for $k=k'=0$ we reproduce
the answer for $\langle\Xi_m|\Xi_m\rangle$ given in \refb{ehk6}.
This gives
\be \label{ehk14a}
S^{(2)} \simeq {1\over 2 
} \langle \Psi_g | \QQ|\Psi_g\rangle
(2\pi)^{26} \, K \, \ln 2 \,
\int d^{26} k  (k^2-1) T(k) T(-k)
\, .
\ee
If we rescale the tachyon fluctuation field 
\be \label{ehk15}
\wh T(k) = \bigg(K \, \ln
2 \, \la\Psi_g |\QQ|\Psi_g\ra \bigg)^{1/2} \,
T(k)\,
,
\ee
then we obtain a canonical kinetic term  
\be \label{ehk16}
S^{(2)} \simeq {1\over 2} \, (2\pi)^{26} \,
\int d^{26} k  (k^2-1) \wh T(k) \wh T(-k)
\, .
\ee

Next we turn to the cubic term in eq.\refb{ehk14}. 
This time we compute  
\ben \label{ehk17}
&& \la \chi_{T}(k_1)|  \chi_{T}(k_2) *
\chi_{T}(k_3)\ra \cr \cr 
&=& K n^{2(k_1^2 + k_2^2 + k_3^2)} \la
e^{ik_1\cdot
X(0)}
e^{ik_2\cdot X((n-1)\pi/2)} e^{i k_3\cdot X(2(n-1)\pi / 2)}
\ra_{\CC_{3n-3}} \nonumber \\
&=& K n^{2(k_1^2 + k_2^2 + k_3^2)} \bigg({4\over 3
(n-1)}\bigg)^{k_1^2+k_2^2+k_3^2}
\la e^{ik_1\cdot
X(1)}
e^{ik_2\cdot X(e^{2\pi i/3})} e^{i k_3\cdot X(e^{4\pi i/3})}
\ra_D \nonumber \\
&\simeq& K\, \bigg({4\over 3\sqrt 3}\bigg)^3\, n^{k_1^2 + k_2^2 +
k_3^2} \, (2\pi)^{26} \delta(k_1+k_2+k_3),
\een
for $k_i^2\simeq 1$ and $n$ large.
The $(\sqrt 3)^{-3}$ factor in the last step came from the correlator on
the disk, computed using a simple generalization of eq.\refb{egenfor}.  
Thus the
cubic term in \refb{ehk14} 
now takes the following form for near on-shell
momenta:
\be \label{ehk18}
-{K\over 3 
}\, (2\pi)^{26} \, \la\Psi_g |\QQ|\Psi_g\ra \,
\bigg({4\over
3\sqrt 3}\bigg)^3 \,
\int d^{26}k_1 d^{26}k_2 d^{26}k_3
\delta(k_1+k_2+k_3) \, T(k_1) T(k_2) T(k_3)\,.
\ee
Expressed in terms of $\wh T$, this reduces to:
\be \label{ehk19}
-{1\over 3} \bigg(K \, 
\la\Psi_g |\QQ|\Psi_g\ra
\bigg)^{-{1\over2}} \hskip-2pt (2\pi)^{26} \,
\bigg({4\over 3\sqrt {3\ln 2}}\bigg)^3 \,
\hskip-6pt\int d^{26}k_1 d^{26}k_2 d^{26}k_3
\delta(k_1+k_2+k_3) \, \wh T(k_1) \wh T(k_2) \wh T(k_3)\,.
\ee
{}From this we see that the on-shell three tachyon coupling is given by:
\be \label{ehk20}
g_T = \bigg(K \, 
\la\Psi_g |\QQ|\Psi_g\ra
\bigg)^{-1/2}
\bigg({4\over 3\sqrt {3\ln 2}}\bigg)^3\, .
\ee
This, in turn, is related to the tension of the D25-brane via the
relation~\cite{9911116,0009191}:
\be \label{ehk21}
\TT_{25} = {1\over 2\pi^2 g_T^2}
= {1\over 2\pi^2} \, {K 
} \, \la\Psi_g |\QQ|\Psi_g\ra \, 
\bigg({ 3\sqrt {3\ln 2} \over 4}\bigg)^{6} \, .  
\ee
On the other hand, the energy density $\EE_c$ associated with the solution
\refb{ehk3} with $|\Psi_m\ra=|\Xi_m\ra$ is given by
\be \label{ehk22}
\EE_c = {1\over 6 
} \la\Psi_g |\QQ|\Psi_g\ra {\la \Xi_m|\Xi_m\ra \over 
V^{(26)}} =
{K\over 6 
}
\, \la\Psi_g
|\QQ|\Psi_g\ra\, .
\ee
{}From eqs.\refb{ehk21} and \refb{ehk22} we get
\be \label{ehk23}
\left( {\EE_c \over \TT_{25}} \right)_{exact}
= {\pi^2 \over 3} \bigg( {16\over 27 \ln 2}\bigg)^3
\simeq 2.0558\,.
\ee
In \cite{0111034}, this ratio
was determined by a direct level truncation computation 
for various values of the level $L$ (table 3(a) of  \cite{0111034}).
A large $L$ extrapolation of their results gives\footnote{
We extrapolated the quantity $H$ to infinite
level with a fit of the form $a_0 + a_1/\log(L)$, then
computed (\ref{ghgh}) using equ.(5.19) of \cite{0108150}.}
\be
\label{ghgh}
\left( {\EE_c \over \TT_{25}} \right)_{numeric} \cong 2.0532 \, ,
\ee 
in very good agreement with our exact prediction (\ref{ehk23}).
The result is of course different from the expected answer of one.
It is also different, though not very much so, from the value of
two, which would suggest that the sliver describes a state with
two D25 branes.

\subsection{Failure of the computation}

We shall now show that
$\T$ fails to satisfy the equations of
motion \refb{ehk13} when we take the inner product of this with the state
$\Tm$.
In other words, we want to show that
\be \label{eleomsl}
\la \Tm |\T\ra \ne \la \Tm| (\T*\Xi_m +\Xi_m * \T)\ra \,.
\ee
We recall from \refb{eorig}  that 
\be \label{eoriggg} 
\la  \chi_{T}(k') |  \, \T \ra  
= n^{2k^2} \,
(2\pi)^{26} \, K\, \delta(k+k') \, .
\ee
%
On the other hand,
\ben  \label{eprod}  
&& \la \Tm | (\Xi_m * \T +\T * \Xi_m) \ra \nonumber \\ \cr
&=& K\,
n^{2 (k^2+ k^{\prime 2})}  \Big[ \la      
e^{i k\cdot X((n-1) \pi/2)}   
e^{i k'\cdot X(0)} \ra_{{\cal
C}_{3n-3} } +  \la      e^{i k\cdot X((n-1) \pi)}   e^{i k' \cdot X(0)}
\ra_{{\cal C}_{3n-3} } \Big] \nonumber \\
&=& K\,
n^{2 (k^2+ k^{\prime 2})} \,  
\bigg({4\over 3(n-1)}\bigg)^{k^2+ k^{\prime 2}} \Big[ \la e^{i k\cdot
X(e^{2\pi i/3})} e^{i k'\cdot X(1)}
\ra_{D} + \la e^{i k\cdot X(e^{4\pi i/3})}    e^{i k'\cdot
X(1)} \ra_{D}\Big] \nonumber \\
&=& 2 \, n^{2k^2} \, (4/3)^{2k^2} (\sqrt{3})^{-2 k^2}\, K \, (2\pi)^{26}
\, \delta(k+k')\, .
\een
Clearly \refb{eoriggg} and \refb{eprod} are not equal for $k^2=1$. This
establishes \refb{eleomsl} and the failure of \refb{ehk14}.

It is useful to note that the sliver itself does satisfy its projector
equation of motion not only against Fock space states, but also 
against the sliver. In
particular if we take
$k=0$, in which case
$\chi_T(k)$ reduces to the sliver, we see from our equations
\refb{T=}  and \refb{Tsliver=}  that $(\Xi * \Xi +
\Xi  * \Xi) - 2 \,\Xi$ has vanishing inner product with the Fock space state.
On  the other hand, eqs. \refb{eoriggg}  and \refb{eprod} show
that
$(\Xi * \Xi + \Xi * \Xi)  - 2\,\Xi$ also has vanishing inner product with the
sliver.\footnote{Since for the matter part alone the normalization of the 
sliver involves infinite factors, in order to make a precise statement one 
needs to work with the sliver of the combined matter ghost CFT. 
In this case the sliver $\Xi$, normalized so that $\la\Xi | Y(i) | 
\Xi\ra=1$, with $Y(i)={1\over 2}c\p c \p^2 c(i)$, satisfies the condition 
that 
$\la\phi|Y(i) | (\Xi - \Xi * \Xi)\ra =0$ when $|\phi\ra$ is a Fock space 
state
of ghost number 0, and also when $|\phi\ra$ is the sliver $|\Xi\ra$.
On the other hand, if we define the state $|\T\ra$ as in eq.\refb{T} with 
$\la \cdot \ra_{\Cn}$ now denoting the correlation function in the combined 
matter ghost CFT, then we shall find that $\la \phi| Y(i) (|\Xi_m * \T\ra + 
|\T * \Xi_m\ra - 2^{k^2-1} |\T\ra)$ vanishes for $k^2=1$ as long as 
$|\phi\ra$ is a Fock space state, but that it does not vanish if $|\phi\ra$ 
is of the form $|\Tm\ra$.}

\sectiono{Discussion}

Given the failure of the HK tachyon state to satisfy a 
strong form of the equation of motion, as indicated in
eq.\refb{eleomsl}, we
see that the computation of the
$\EE_c/\TT_{25}$ ratio outlined in refs.\cite{0108150,0111034} is not
valid. This leaves open the question as to whether the identification of
the HK
state with the
tachyon fluctuation around the D25-brane background could still
be valid. The tachyon certainly does satisfy
the linearized equation of motion around the D25-brane solution for
$k^2=1$ in a weak sense, {\it i.e.} when we consider the inner product
of the equations of motion with a Fock space state. On the other hand
it fails to
satisfy the equations of motion when we consider its inner product with
the tachyon state itself. Clearly a more satisfactory construction of the
tachyon fluctuation will be one for which the inner product of the
equations of motion with any state yields the correct on-shell condition
for the tachyon.

We note 
that the problem discussed here is a special 
case of a more general ambiguity 
that arises in dealing with correlation functions of states of the type 
\refb{eHK}.
In particular, 
when computing various correlation 
functions we have regularized the sliver by replacing it by a finite $n$ 
wedge state $|n\ra$, and {\it have chosen the same value of $n$ for the
sliver $|\T\ra$ and $|\chi_T(k')\ra$.} If we had chosen different 
values of $n$ for regularizing different states, we would have gotten 
different answers for these 
correlators. Thus in general we need to be very careful in dealing with 
such states.

One could of course compute these correlation functions also 
using oscillator methods as in ref.\cite{0108150}. One might wonder if the 
oscillator computation suffers from similar ambiguities. To this end we 
note that
the expressions involving these correlation  functions 
involve inverse powers of the matrix $(1+3 M^{11})$, where 
$M^{rs}=CV^{rs}$, with $C_{mn}=(-1)^m \delta_{mn}$ 
the charge 
conjugation matrix, and $V^{rs}_{mn}$ 
the matter Neumann 
coefficients. Due to the presence of a $(-1/3)$ eigenvalue of the matrix 
$M^{11}$~\cite{0111034,0111069,rsz}, the matrix $(1+3 M^{11})$ is actually 
singular, and as a result various formal identities break down in the 
presence of inverse powers of $(1+3M^{11})$. In particular, although 
formally $[M^{rs}, M^{r's'}]$ vanishes, it turns out that 
$[M^{rs},M^{r's'}](1+3 M^{11})^{\alpha}$ does not vanish 
in general for $\alpha<0$, and definitely does not vanish
for $\alpha\leq -3/2$. 
Thus different expressions which are formally identical have  different
values in practice. This, in turn, makes the correlation  functions
involving states of type
\refb{eHK} ambiguous.
The eigenvalue spectrum of the matrix $M^{11}$ will be discussed in
a  separate paper~\cite{rsz}. 

An alternative proposal for computing the ratio $\EE_c/\TT_{25}$
was given in \cite{0111034} in which it was noted that the 
expression for the ratio $\EE_c/\TT_{25}$ in the oscillator approach 
contains a 
term which is formally zero, but because of the presence of inverse 
powers of the matrix $(1+3 M^{11})$, it does not vanish. In analysing 
this term, the authors 
expand the expression for $\EE_c/\TT_{25}$ in a
power  series expansion in $(1+3 M_{11})$ and for those terms containing
powers 
larger than $-3/2$, 
simplify the expressions by
using  the formal identities satisfied by the matrices $M^{rs}$. This gives
a  simpler expression for $\EE_c/\TT_{25}$. The numerical result for this 
modified expression also comes out to be close to two. The precise 
correspondence between this modified expression and the conformal field 
theory description is not entirely clear to us at present.

\bigskip
\noindent{\bf Acknowledgements}:
We are grateful to H.~Hata and S.~Moriyama for 
a stimulating correspondance. We would also like to
acknowledge discussions with   
D.~Gaiotto,  T.~Kawano and W.~Taylor.
The work of L.R. was supported in part
by Princeton University
``Dicke Fellowship'' and by NSF grant 9802484.
The  research of A.S. was supported in part by a grant 
from the Eberly College 
of Science of the Penn State University.
The work of  B.Z. was supported in part
by DOE contract \#DE-FC02-94ER40818.

\end{document}